\documentclass[prb,preprint,showpacs,preprintnumbers,amsmath,amssymb]{revtex4}

\usepackage{graphicx}
\usepackage{dcolumn}
\usepackage{bm}
 
\begin{document}
\newcommand{\braopket}[3]{\langle #1 | \hat #2 |#3\rangle}
\newcommand{\braket}[2]{\langle #1|#2\rangle}
\newcommand{\bra}[1]{\langle #1|}
\newcommand{\braketbraket}[4]{\langle #1|#2\rangle\langle #3|#4\rangle}
\newcommand{\braop}[2]{\langle #1| \hat #2}
\newcommand{\ket}[1]{|#1 \rangle}
\newcommand{\ketbra}[2]{|#1\rangle \langle #2|}
\newcommand{\op}[1]{\hat {#1}}
\newcommand{\opket}[2]{\hat #1 | #2 \rangle}

\title{Non-collinear coupling between magnetic adatoms in carbon nanotubes}

\author{A. T. Costa}
\email{antc@if.uff.br}
\affiliation{
Instituto de F\'isica, Universidade Federal Fluminense, 24210-346 Niter\'oi, RJ, Brazil
}
\author{C. G. Rocha}
\affiliation{
School of Physics, Trinity College Dublin, Dublin 2, Ireland
}
\author{M. S. Ferreira}
\email{ferreirm@tcd.ie}
\affiliation{
School of Physics, Trinity College Dublin, Dublin 2, Ireland
}

\date{\today}

\begin{abstract}

The long range character of the exchange coupling between localized magnetic
moments indirectly mediated by the conduction electrons of metallic hosts often
plays a significant role in determining the magnetic order of low-dimensional
structures. In addition to this indirect coupling, here we show that the direct
exchange interaction that arises when the moments are not too far apart may
induce a non-collinear magnetic order that cannot be characterized by a
Heisenberg-like interaction between the magnetic moments. We argue that this
effect can be manipulated to control the magnetization alignment of magnetic
dimers adsorbed to the walls of carbon nanotubes.

\end{abstract}

\maketitle
\bibliographystyle{apsrev} 

More than a decade after the discovery of carbon nanotubes, these nanoscale
cylindrical structures are still the subject of intensive scientific research
due to their intriguing physical properties. Significant progress has been made
to explain the intrinsic properties of nanotubes but to expand their
applicability we must understand and control how they interact with other
objects. Nanotubes interacting with magnetic foreign objects offer a wide range
of technologically promising possibilities within the so-called area of
spintronics in molecular structures. In fact, the ability to produce sizeable
changes in the conductance of a nanotube due to an applied magnetic field is one
of the driving forces in the research of magnetic properties of carbon-based
structures \cite{alphenaar, littlewood}.  Substrates\cite{ferreira04, cespedes04},
substitutional impurities\cite{fazzio1}, adsorbed atoms\cite{fazzio2, fazzio3}
and nanoparticles\cite{yang03} are some of the different magnetic foreign
objects that can interact with carbon nanotubes. Among those, transition-metal
magnetic adatoms have been reported to produce noticeable changes in the
spin-dependent electronic structure of carbon nanotubes.\cite{yang03,fazzio2}
Furthermore, the formation of defect-induced magnetic moments in carbon-based
materials appears as an additional possibility to manipulate the magnetic
response of these systems\cite{krashenninikov}.

The transport properties of magnetically-doped structures depend on the way
their magnetic moments are oriented. Therefore, besides establishing how
magnetic impurities affect the electronic structure of a nanotube, it is crucial
to understand the nature of the coupling between these moments. Regarding
magnetic adatoms, dipolar and exchange interactions are the basic mechanisms
defining the alignment of their moments \cite{ze+dora}. The former decays rather
quickly as the moments are moved apart whereas the latter depends on both the
dimensionality and on the nature of the interaction. Direct exchange coupling
results from the overlap between wave functions centred at the magnetic
impurities but also decays abruptly as the distance between them increases. Of
indirect nature, the so-called indirect exchange coupling (IEC) between magnetic
impurities mediated by the conduction electrons of the non-magnetic host is
known to decay more slowly and plays an important role in determining the
overall magnetic alignment of the system. In fact, we have recently shown that
the IEC between magnetic adatoms in carbon nanotubes is long ranged and is of
Heisenberg form \cite{coupling1}. We have also shown that the coupling may be
either ferromagnetic or antiferromagnetic, depending on the type of atoms, their
separation as well as on the size and type of nanotube host.

The slow decaying rate of the IEC makes the indirect interaction dominant when
magnetic objects are sufficiently far apart. However, it becomes comparable to
the direct exchange coupling when the objects are brought closer together.
In this situation, the proximity of two magnetic moments may give rise to
unusual ordering. Recent calculations indicate that the magnetization
alignment of magnetic dimers adsorbed onto the surface of a 3-dimensional metal
is often neither ferromagnetic nor antiferromagnetic but follows a non-collinear
order \cite{bola-prl2005}. The origin of such a non-Heisenberg behaviour lies in
the competition between the direct and indirect contributions to the exchange
coupling and should arise when localized magnetic moments embedded in a metallic
host are in close proximity. It is the purpose of this manuscript to show that
such an effect also occurs when magnetic dimers are adsorbed onto the walls of a
metallic nanotube. Since the IEC decays more slowly in low-dimensional hosts, we
argue that nanotubes are likely to expand the range of separations for which
this competition can induce non-collinear alignments. In other words, the effect
of non-collinear magnetization alignment of dimers should be more robust in
nanotubes than in flat substrates of higher dimensionality.

\begin{figure}
\begin{tabular}{cc}
\includegraphics[width = 0.6\textwidth]{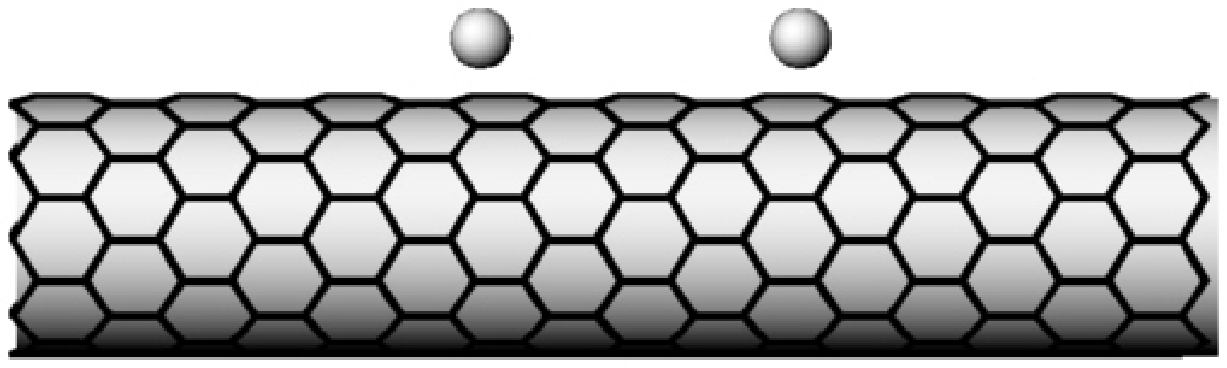} &
\includegraphics[width = 0.35\textwidth]{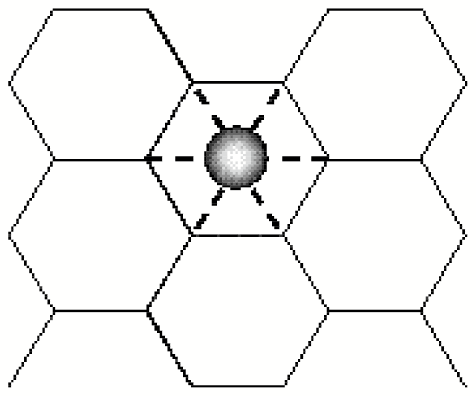}
\end{tabular}
\caption{Left: Schematic diagram representing a magnetic dimer formed by two 
magnetic atoms adsorbed to the surface of a carbon nanotube. Right: A more 
detailed diagram shows a 2-dimensional depiction of the magnetic adatom above 
the hexagonal lattice of the nanotube. The dashed lines highlight to which 
nearest-neighbour sites the adatom is connected by a hopping parameter $t$.}
\label{figure_1}
\end{figure}

We consider a dimer made of two magnetic atoms, labelled $A$ and $B$, adsorbed
to the walls of an infinitely long carbon nanotube and schematically represented
in Figure \ref{figure_1}. Magnetism in these atoms is driven by an intra-atomic
Coulomb interaction that, when treated in mean-field approximation through a
self-consistent procedure, can be described by an effective spin-dependent
potential located at the atomic positions. In this way, the electronic structure
of the entire system is well described by a Hamiltonian in a basis of localized
atomic orbitals. In such a basis, this tight-binding-like Hamiltonian is fully
determined by the on-site potentials and hopping integrals.

We assume that the magnetic moments of the individual adatoms are initially
parallel, hereafter referred to as the ferromagnetic configuration. In this
configuration, the Hamiltonian of the entire system written in the basis $| j
\rangle$ of atomic orbitals centred at a site $j$ is given by $\hat{H} =
\hat{H}_{NT} + \hat{H}_D + \hat{V}_C$, where
$\hat{H}_{NT}=\sum_{j,j^\prime}|j\rangle\gamma\langle j^\prime |$ is the
Hamiltonian of the individual nanotube, $\hat{H}_D =|a\rangle\epsilon_a\langle
a | + |b\rangle\epsilon_b\langle b | + |a\rangle \tau \langle b | + |b\rangle
\tau^\dag \langle a | $ is the Hamiltonian associated with the dimer and
$\hat{V}_C = \sum_{\ell} t\left\{|a\rangle \langle \ell | + |\ell\rangle
\langle a |\right\} + \sum_{\ell^\prime} t^\dag \left\{|b\rangle \langle
\ell^\prime | + |\ell^\prime\rangle  \langle b |\right\}$ refers to the
coupling between the adatoms and the nanotube. The parameters $\gamma$,
$\epsilon_a$, $\epsilon_b$ and $t$ are all matrices in spin and orbital spaces
and correspond to the hopping between nearest-neighbour sites in the nanotube,
the on-site potentials of atom A, of atom B and the hopping between the nanotube
atoms and the adatoms, respectively. Likewise, the basis $\vert j \rangle$
represents vectors in the same linear space. It is evident from the expressions
above that sites $j=a$ and $j=b$ label the two adatoms and the indices $\ell$
and $\ell^\prime$ label the nanotube sites that are coupled to those respective
atoms. These latter sites are schematically illustrated in Figure
\ref{figure_1}b where a magnetic adatom lying just above the centre of the
nearest hexagon is connected to its nearest carbon atoms. It is worth mentioning
that there are two fundamental differences between $\hat{H}$ and
the Hamiltonian considered in Reference [\onlinecite{coupling1}]. The
inclusion of a hopping $\tau$ between atoms A and B that accounts for the
direct interaction due to their proximity is the first difference, followed by
the fact that the magnetic adatoms are now located above the centres of the
nanotube hexagons, as opposed to being immediately above the carbon atoms.

The energy required to rotate the magnetic moment of one adatom relatively to
the other is given by\cite{coupling1}
\begin{equation}
\Delta {\cal E}(\theta) = \frac{1}{\pi}\int_{-\infty}^{\infty}d\omega
\,\left[\frac{1}{1 + e^{\beta (\omega-\mu)}}\right] \, {\rm Im}\,{\rm
Tr} \, \ln [1 + 2 \, V_x^2 \, (1 - \cos\theta)
G_{a,b}^{\uparrow}(\omega)\, G_{b,a}^{\downarrow}(\omega)\,]\, ,
\label{spin-indices}
\end{equation}
where $\theta$ is the angle of rotation and $V_x$ is a matrix in orbital space
representing the strength of the local exchange potentials.
$G_{m,\ell}^{\sigma}(\omega)$ represents the propagator between sites $j=\ell$
and $j=m$ for electrons of spin $\sigma$ and energy $\omega$ in the FM
configuration, and the trace is over orbital indices. Regarding the fraction
within brackets, $\beta=1/k_B T$, where $k_B$ is the Boltzmann constant and $T$
is the temperature.

For the sake of simplicity the electronic structure of the system will be here
treated within the single-band tight-binding model. The expressions above are
very general and by no means restricted to such a simple case. The results here
obtained can be easily extended to a multi-orbital description but bring no
qualitative difference. This is justified by the fact that the main features of
the IEC are predominantly determined by the extended electrons of the host, in
this case the nanotube, whose electronic structure is known to be well
reproduced by a single-band tight-binding model. Having transition-metal atoms
in mind, the adatoms are then described by a 5-fold degenerate $d$-band with the
appropriate occupation to represent typical magnetic materials. In this way,
rather than matrices in orbital indices, all quantities in the integrand of Eq.
(\ref{spin-indices}) become scalar.

The energy variation $\Delta {\cal E}(\theta)$ can be calculated by inserting
the appropriate Green function matrix elements $G_{a,b}^{\uparrow}(\omega)$ and
$G_{b,a}^{\downarrow}(\omega)$ into Eq.(\ref{spin-indices}) and evaluating the
corresponding energy integral. The Green functions are evaluated firstly in the
absence of any magnetic impurities through a standard renormalization
procedure \cite{claudia} and subsequently renormalized through Dyson's equation
to account for the presence of the magnetic adatoms in a self-consistent
fashion. As far as the tight-binding parameters are concerned, the on-site
potentials $\epsilon_a$ and $\epsilon_b$ are easily determined by the
aforementioned self-consistent procedure and follow from the appropriate
selection of the d-band occupation. The hopping parameter $\gamma = 2.7 \, {\rm eV}$ is known to describe well the $\pi$-band of carbon nanotubes. The only
remaining parameters to be specified are the coupling $t$ describing the
electronic hopping linking the adatoms with the nanotube and the proximity
parameter $\tau$ that accounts for the direct interaction between nearby
adatoms.

\begin{figure}
\begin{tabular}{c}
\includegraphics[width=0.8\textwidth,clip]{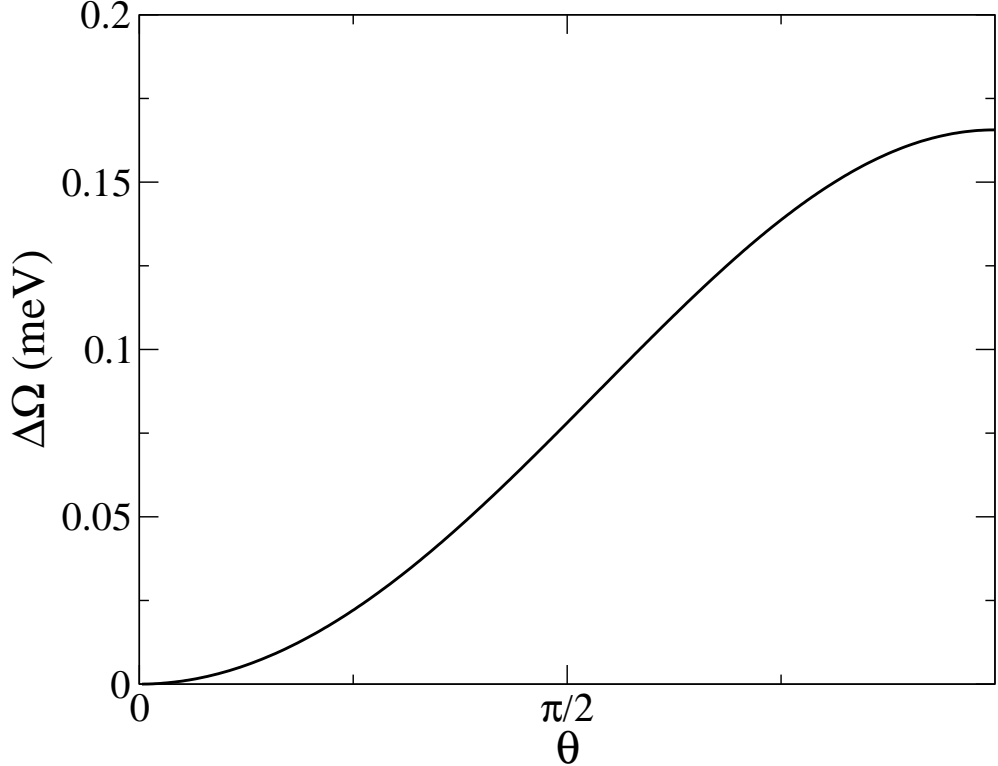}
\end{tabular}
\caption{Energy change $\Delta\Omega$ as a function of relative angle $\theta$
between magnetizations for two Mn atoms ($n=1.1$ electron per atom). The hopping between
magnetic atoms has been artificially turned off ($\tau=0$), and the hopping Mn-C
is $t=0.4\gamma$. Each Mn atom is atop the centre of a hexagon that are 4 unit cells apart. }
\label{coupling_tau0}
\end{figure}

We start by assuming $\tau = 0$, which corresponds to the standard IEC between
magnetic objects without any direct interaction\cite{coupling1}. In Fig.
\ref{coupling_tau0} we see that the IEC has a nice sinusoidal behavior as a
function of the angle between magnetizations if the adatoms are sufficiently far apart. As demonstrated many years
ago by Herring \cite{herring}, the angular dependence of the energy variation
follows a standard Heisenberg form $\Delta {\cal E}(\theta) = J_0+J_I
\cos(\theta)$ only for large separations of the magnetic objects. The deviation
from the traditionally assumed Heisenberg-like behaviour may arise as a consequence of
charge rearrangements due to the relative rotation of magnetic moments. Such
rearrangements are of little importance in a 3D metal because screening is
extremely efficient there; such is not the case in carbon nanotubes, even
metallic ones. In fact, for small separations one finds small deviations from the Heisenberg form which are nevertheless not strong enough to introduce any non-collinearity in the equilibrium states.  For the Heisenberg-like case it is simple to show that the
sinusoidal behaviour remains the same for arbitrary values of $t$ and that the
only effect that this parameter has on $\Delta {\cal E}(\theta)$ is in
determining the amplitude of the cosine function, that is, on the quantity
$J_I$. 

Whereas ab-initio calculations would struggle to provide a detailed
angular dependence of the total energy, they can be used to obtain the actual
total energy of the system in both ferromagnetic and antiferromagnetic
configurations, two quantities whose difference provides direct information on
$\Delta {\cal E}(\theta)$ and, in turn, on the reasonable range of values for
the parameter $t$. With that purpose in mind, ab-initio calculations were
evaluated for adatoms lying a minimum distance apart to reflect the absence of
any direct interaction. The difference between the total energy values for the
antiferromagnetic and ferromagnetic configurations was then used to guide us to
the appropriate range of values of the hopping $t$. Our calculations are
performed using the SIESTA code\cite{siesta} within the LDA and GGA approaches 
for the exchange and correlation potential\cite{lda}. Norm-conserving
pseudopotentials\cite{pseudo} with relativistic corrections and a split-valence
double-$\zeta$ basis of pseudoatomic orbitals with an orbital confining energy
of 0.05 eV and an energy cutoff of 150 Ry were used. The $k$-point sampling is composed of
11 $k$ points in the axial direction of the tube and we built an unit cell with
96 atoms for the (4,4) CN plus the two magnetic atoms. One may notice
that we induced a magnetization up or down only in the magnetic atoms. The two
configurations (ferro and antiferromagnetic) for the adatoms were studied. The
in-plane lattice parameter was chosen to be large enough to guarantee a
negligible interaction between periodic CN images ($\sim 98$\AA). All the
structure was optimized by conjugate gradient with force tolerance of 0.05
eV/\AA. 

\begin{figure}
\centerline{\includegraphics[width=0.8\textwidth]{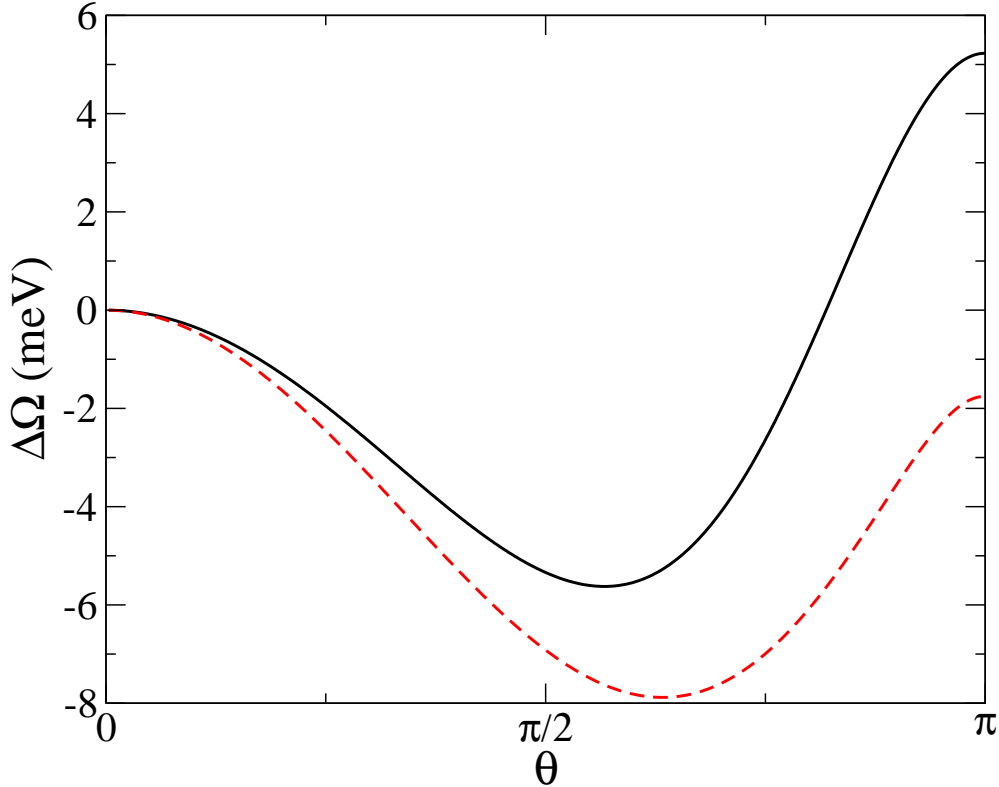}}
\caption{(Color online) Energy change as a function of angle between magnetizations of two Mn
atoms adsorbed to a (5,5) nanotube. The adatoms are located atop the centres of
nearest neighbour hexagons, connected to six C atoms each. The Mn-Mn hopping
$\tau=0.47\gamma$. The dashed curve
corresponds to a doped nanotube, simulated by a change in Fermi energy
$\Delta E_f = 0.2\gamma$.}
\label{non_collinear}
\end{figure}

Having determined the range of values of the parameter $t$ that accounts for the
correct magnitude (and sign) of the indirect coupling $J_I$, we can now proceed
to introduce a direct hopping term $\tau$ that corresponds to bringing the two
adatoms closer together. As depicted in Figure \ref{non_collinear}, a suitable choice of values for the parameter $\tau$ may induce an interesting situation in which the angular dependence of the
energy variation is no longer described by a single sinusoidal function. Such a
non-Heisenberg-like behaviour arises as the result of a competition between the
direct and indirect contributions to the magnetic coupling. Having the energy
minimum at an intermediate angle $0 < \theta < \pi$ means that the preferential
magnetization alignment for such a magnetic dimer is of non-collinear type.
Incidentally, one may notice that the effect is robust under doping or
the application of a gate voltage: the dashed curve in figure \ref{non_collinear}
shows that the stable configuration is still non-collinear after a change of
$0.2\gamma$ in the Fermi energy. It is interesting to notice that, upon doping,
the energy of the antiferromagnetic state becomes smaller than that of the
ferromagnetic alignment. First principle calculations that only evaluate total
energies or energy differences between the two collinear states would predict a
change in alignment upon doping. What our results show is that, for magnetic
objects close to each other, those energy differences are not enough: one must
analyze the local stability around the collinear states.

\begin{figure}
\begin{tabular}{c}
\includegraphics[width=0.8\textwidth,clip]{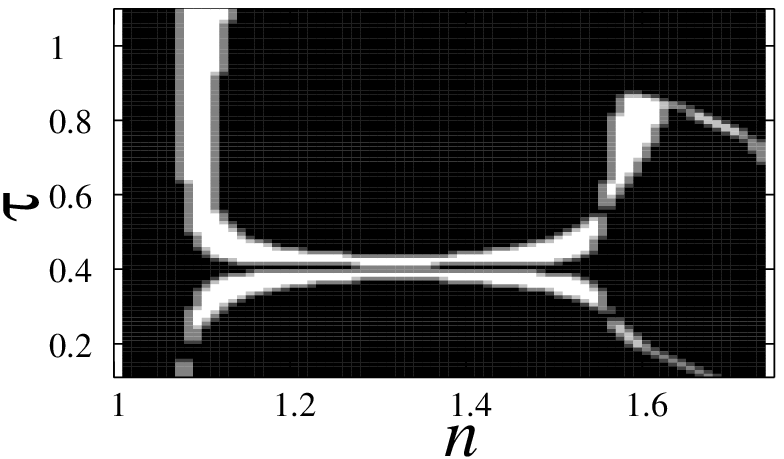}\\
\includegraphics[width=0.8\textwidth,clip]{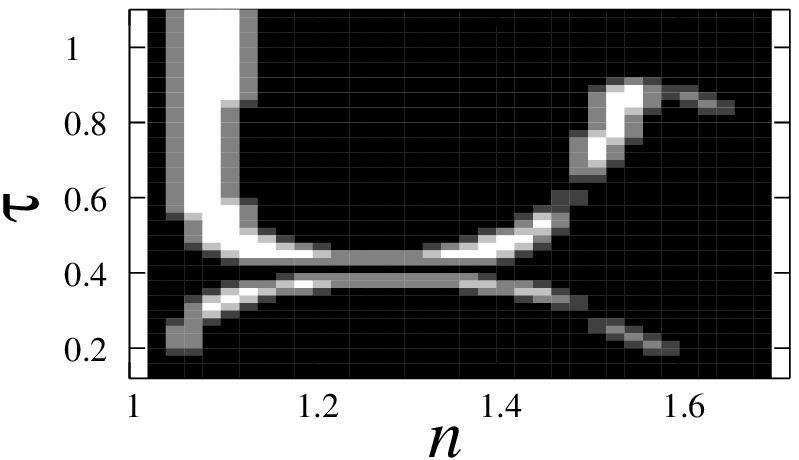}
\end{tabular}
\caption{Top panel: state diagram for the ground state of a magnetic dimer 
on an armchair (5,5) nanotube, and $t=0.7$. The white region corresponds to 
non-collinear ground states. Bottom panel: same as above, but the Fermi energy
has been shifted by $0.2\gamma$.}
\label{diagram_tl0.7}
\end{figure}

Rather than the result of a fortunate choice of parameters, this
non-collinearity effect is robust enough to be seen in a wide range of values
for the proximity parameter $\tau$ as well as for adatoms of different nature.
To assess the robustness of this effect we introduce a parameter $\zeta$ defined
as
\begin{equation} 
\zeta = SGN \left[ \frac{ 
\left( \frac{d^2\Delta{\cal E}}{d\theta^2}\right)_{\theta=0} 
\times \left(\frac{d^2\Delta{\cal E}}{d\theta^2} \right)_{\theta=\pi} } 
{ \left\vert \left( \frac{d^2\Delta{\cal E}}{d\theta^2} \right)_{\theta=0} 
\times \left( \frac{d^2\Delta {\cal E}}{d\theta^2}  \right)_{\theta=\pi} 
\right\vert  } \right]
\label{zeta}
\end{equation}
where the function $SGN(x) = 1$ for $x>=0$ and $SGN(x) = -1$ for $x<0$.
Eq.(\ref{zeta}) uses the curvature of the function $\Delta {\cal E}(\theta)$ at
$\theta = 0$ and $\theta=\pi$ to define an indirect way of testing whether or
not the magnetic coupling of the dimer follows a Heisenberg law. According to
the definition above, $\zeta = -1$ corresponds to a Heisenberg behaviour and
$\zeta = +1$ corresponds to the non-collinearity effect. A more informative view
is in the diagram of Figure \ref{diagram_tl0.7} where we have adopted a colour code that
uses black to represent $\zeta=+1$ and white to represent $\zeta=-1$. On the
horizontal axis the proximity parameter $\tau$ runs from $10\%$ to $110\%$ of the
electronic hopping $\gamma$, whereas the d-band occupation on the vertical axis
includes characteristic values corresponding to Fe and Co, for instance. It is
noteworthy that the non-collinear state takes a non-negligible fraction of the
diagram, indicating that the situation illustrated in Figure \ref{non_collinear} is far from
coincidental.

Since this effect results from a competition between the direct and indirect
contributions to the magnetic coupling, it is natural to suspect that one may
have the ability to select the preferred alignment of the magnetizations by
controlling how these two contributions are related. Although changing $\tau$ is
a mathematically possible way of altering the direct contribution to the
coupling, it is of little applicability once this would, in practical terms,
involve varying the separation of the magnetic atoms in the dimer. A possible
alternative is to change the indirect contribution to the coupling. To change
the indirect coupling by moving the adatoms apart would be equally impractical
but one could make use of the fact that this type of coupling can also be
affected by controlling the nanotube Fermi energy. In fact, the authors have
already shown\cite{coupling1} that the indirect coupling can be modified by
tuning the Fermi energy through doping or through a carefully engineered gate
voltage. The bottom panel of Figure \ref{diagram_tl0.7} shows a similar diagram 
in which the Fermi level of the nanotube host has been shifted by $\Delta E_F = 0.2\gamma$. 
A clear distinction between the diagrams in top and bottom panels of 
Figure \ref{diagram_tl0.7} corroborates that this is in fact a
possible and more practical way of controlling the magnetization alignment of
magnetic dimers attached to the walls of a carbon nanotube. Whereas a single
isolated dimer may not be able to produce a sizeable effect, a finite
concentration of those dimers will undoubtedly be sufficient to produce a
measurable change in the magnetization of the system.

\begin{figure}
\begin{tabular}{c}
\includegraphics[width=0.8\textwidth,clip]{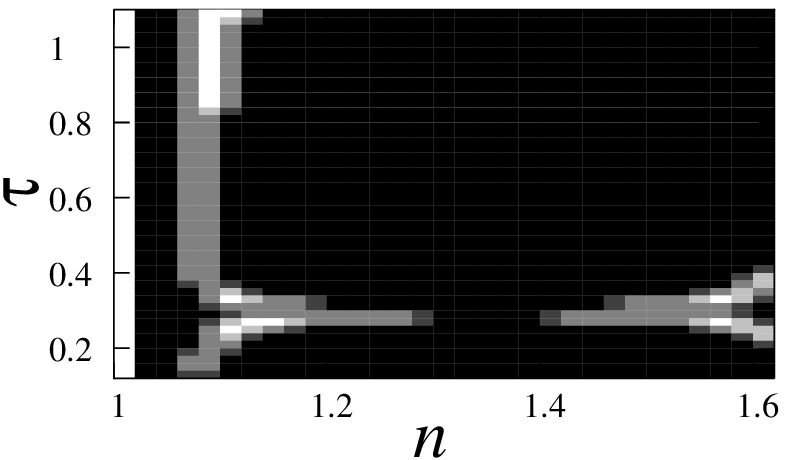} \\
\includegraphics[width=0.8\textwidth,clip]{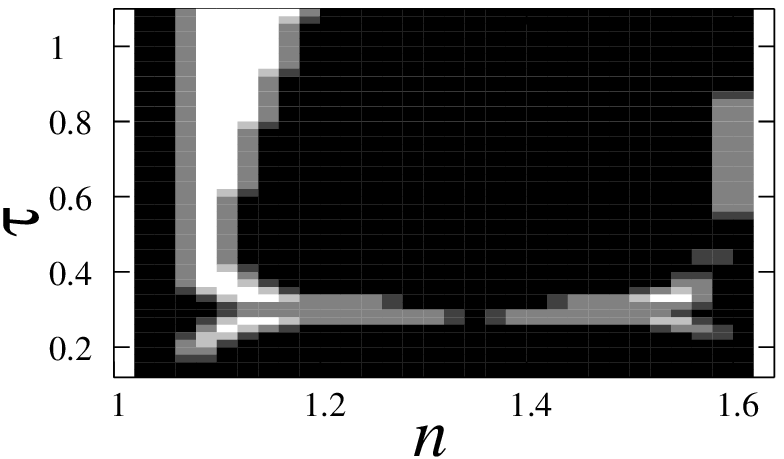}
\end{tabular}
\caption{State diagram for the ground state of a magnetic dimer on an 
armchair (5,5) nanotube, and $t=0.7\gamma$, for two values of the effective
Coulomb interaction: $U=20\gamma$ (top panel) and $U=12\gamma$ (bottom panel). 
The white region corresponds to non-collinear ground states.}
\label{diagrams_tl0.6_U20_12}
\end{figure}

In our model calculations we must choose a value for the effective Coulomb interaction
parameter $U$ such that the Stoner criterion for the onset of magnetism is satisfied \cite{yosida}.
In single-orbital models this value has to be unrealistically large due to the artificially reduced density
of states at the Fermi level. It is well known that the large $U$ limit of the Hubbard model is
the Heisenberg model \cite{yosida}. Since we are looking for deviations from Heisenberg-like
behavior, the large $U$ constraint may force us to underestimate the relative frequency
of occurrence of non-collinearity. It is, thus, instructive to examine the behavior of the
kind of state diagram presented in figure \ref{diagram_tl0.7} under variations of the
Coulomb parameter $U$. In figure \ref{diagrams_tl0.6_U20_12} it is obvious that a
smaller $U$ results in a significantly expanded region of non-collinear ground-states. It is
important to notice, however, that even for the high $U=20\gamma$ we used to obtain
the results in figure \ref{diagram_tl0.7} non-collinear ground-states occupy a significant part
of the state diagram.

In summary, we have shown that the proximity of two magnetic adatoms attached to
the walls of carbon nanotubes may induce the formation of non-collinear
alignment of their magnetizations. This effect is the result of a competition
between the direct and indirect contributions to the exchange coupling, which
become comparable when the magnetic adatoms are not too far apart. Due to the
long-range character of the IEC in low-dimensional structures, carbon nanotubes
are ideal candidates to display the non-heisenberg behaviour discussed in this
paper. Moreover, the ability to control the indirect coupling through a careful
selection of the Fermi energy of nanotubes opens the road to the possibility of
controlling the magetization of nanotube-based systems doped wih magnetic
dimers.

\begin{acknowledgments}
The authors gratefully acknowledge many enlightening conversations with
R. B. Muniz. M. S. F. acknowledges the financial support of Science
Foundation Ireland and Enterprise Ireland. A. T. C acknowledges financial
support received from CNPq and FAPEMIG from Brazil.
\end{acknowledgments}

\end{document}